\documentclass[paper]{JHEP3}
\usepackage{epsfig}
\usepackage{graphicx}
\usepackage{xspace}
\usepackage{amssymb}
\usepackage{subfigure}
\usepackage{multirow}

\newcommand{\ra} {\rightarrow}

\newcommand{\be}{\begin{equation}}
\newcommand{\ee}{\end{equation}}

\newcommand{\bea}{\begin{eqnarray}}
\newcommand{\eea}{\end{eqnarray}}
\newcommand{\beanon}{\begin{eqnarray*}}
\newcommand{\eeanon}{\end{eqnarray*}}
\newcommand{\ba}{\begin{array}}
\newcommand{\ea}{\end{array}}
\newcommand{\bd}{\begin{description}}
\newcommand{\ed}{\end{description}}
\newcommand{\bi}{\begin{itemize}}
\newcommand{\ei}{\end{itemize}}
\newcommand{\ben}{\begin{enumerate}}
\newcommand{\een}{\end{enumerate}}
\newcommand{\bc}{\begin{center}}
\newcommand{\ec}{\end{center}}

\newcommand{\GeV}{\mbox{${\mathrm GeV}$}\xspace}

\newcommand{\ordEW}{\mathcal{O}(\alpha_{\scriptscriptstyle EM}^6)\xspace}
\newcommand{\ordQCD}{\mathcal{O}(\alpha_{\scriptscriptstyle EM}^4
  \alpha_{\scriptscriptstyle S}^2)\xspace}
\newcommand{\ordQCDsq}{\mathcal{O}(\alpha_{\scriptscriptstyle EM}^2
  \alpha_{\scriptscriptstyle S}^4)\xspace}

\newcommand{\eqn}[1]{Eq.(\ref{#1})}

\newcommand{\fig}[1]{Fig.~\ref{#1}}

\newcommand{\sect}[1]{Sect.~\ref{#1}}

\newcommand{\Phantom}{{\tt PHANTOM}\xspace}

\newcommand{\MadEvent}{{\tt MADEVENT}\xspace}

\def\pl #1 #2 #3 {{\it Phys.~Lett.} {\bf#1} (#2) #3}   
\def\np #1 #2 #3 {{\it Nucl.~Phys.} {\bf#1} (#2) #3}
\def\zp #1 #2 #3 {{\it Z.~Phys.} {\bf#1} (#2) #3}
\def\pr #1 #2 #3 {{\it Phys.~Rev.} {\bf#1} (#2) #3}
\def\prep #1 #2 #3 {{\it Phys.~Rep.} {\bf#1} (#2) #3}
\def\prl #1 #2 #3 {{\it Phys.~Rev.~Lett.} {\bf#1} (#2) #3}
\def\intj #1 #2 #3 {{\it Int. J. Mod. Phys.} {\bf#1} (#2) #3}
\def\mpl #1 #2 #3 {{\it Mod.~Phys.~Lett.} {\bf#1} (#2) #3}
\def\rmp #1 #2 #3 {{\it Rev. Mod. Phys.} {\bf#1} (#2) #3}
\def\cpc #1 #2 #3 {{\it Comp. Phys. Commun.} {\bf#1} (#2) #3}
\def\epj #1 #2 #3 {{\it Eur. Phys. J.} {\bf#1} (#2) #3}
\def\jhep #1 #2 #3 {{\it JHEP} {\bf#1} (#2) #3}

\title{Multiple Parton Interactions, top--antitop and $W+4j$ production
at the LHC}

\author{
Ezio Maina$^{a,b}$\\
$^a$ INFN, Sezione di Torino, Italy,\\
$^b$ Dipartimento di Fisica Teorica, Universit\`a di Torino, Italy
}

\preprint{DFTT 36/2009}

\abstract{
The expected rate for Multiple Parton Interactions (MPI) at the LHC is large. This
requires an estimate of their impact on all measurement foreseen at the LHC
while opening unprecendented opportunities for a detailed study of these
phenomena. In this paper we examine the MPI background to top-antitop production,
in the semileptonic channel, in the early phase of data taking when the full
power of $b$--tagging will not be available. The MPI background turns out
to be small but non negligible, of the order of 20\% of the background provided
by $W+4j$ production through a Single Parton Interaction.
We then analyze the possibility of studying Multiple Parton Interactions
in the $W+4j$ channel, a far more complicated setting than the reactions 
examined at lower energies. The MPI contribution turns out to be dominated by
final states with two energetic jets which balance in transverse momentum, and it
appears possible, thanks to the good angular resolution  of ATLAS and CMS,
to separate the Multiple Parton Interactions contribution from
Single Parton Interaction processes.
The large cross section for two jet production suggests that also
Triple Parton Interactions (TPI) could provide a non
negligible contribution. Our preliminary analysis suggests that it might be
indeed possible to investigate TPI at the LHC.
}  

\begin{document}

\section{Introduction}
\label{sec:intro}

The presence of Multiple Parton Interactions (MPI) in high energy hadron
collisions has been convincingly demonstrated 
\cite{Akesson:1986iv,CDF_MPI}.

MPI rates at the LHC are expected to be large,
making it necessary to estimate their contribution to the background of
interesting physics reactions. Previous studies evaluated the MPI background to
Higgs production in the channel $pp\ra WH \ra  l\nu b \bar{b}$,
\cite{DelFabbro:1999tf}, 
4$b$  production \cite{DelFabbro:2002pw} and $WH$, $ZH$ production
\cite{Hussein:2006xr}.  
On the other hand, their abundance at the LHC makes it
possible to study MPI experimentally in details, testing and validating the
models which are used in the Monte Carlo's
\cite{Sjostrand:2004pf,Sjostrand:2004ef,Butterworth:1996zw,Bahr:2008dy}
to describe these important features
of hadron scattering. It is therefore of interest to search for new reactions in
which MPI can be probed and to study in which kinematical regimes they are best
investigated. Different reactions involve different combinations of initial
state partons, for instance $\gamma+3j$, $Z+3j$, $W+3j$ and $4j$ MPI processes
test specific sets of quark and gluon distributions inside the proton.
The comparison of several MPI processes will also allow to study the possible
$x$--dependence of these phenomena, namely the dependence
on the fraction of momentum carried by the partons.

CDF found no evidence of 
$x$--dependence in their data which included jets of transverse momentum as low
as five GeV. However in
Ref.\cite{Snigirev:2003cq,Korotkikh:2004bz,Cattaruzza:2005nu} it was shown that
correlations between the value of the double distribution functions for
different values of the two momentum fractions $x_1, x_2$ are to be expected,
even under the assumption of no correlation at some scale $\mu_0$, as a
consequence of the evolution of the distribution functions 
to a different scale $\mu$, which is determined
by an equation analogous to the usual DGLAP equation.
In \cite{Cattaruzza:2005nu} the corrections to the factorized form for the 
double distribution functions have been estimated. They depend on the
factorization scale, being larger at larger scales $Q$, and on the $x$ range,
again being more important at larger momentum fractions. For $Q = M_W$ and
$x \sim 0.1$ the corrections are about 35\% for the gluon-gluon case.    
Moreover  
Ref.\cite{Cattaruzza:2005nu} showed that the correlations in $x_1, x_2$ space
are different for different pairs of partons, pointing to an unavoidable flavour
dependence of the double distribution functions.

\eject

In this paper we examine 
\bi
\item the background generated by MPI to $t \bar{t}$ production particularly in
the early phase of data taking at the LHC in which $b$--tagging might perform
poorly or not at all;
\item the possibility of studying MPI in the $W+4j$ channel.
\ei
In all casses we assume that one W decays leptonically, allowing for easy
triggering.

The LHC will be a top-antitop factory and the large rate will allow accurate 
measurements of the top mass and cross section. The top mass is one the
most crucial ingredients in the high precision global fit of the Standard Model
and together with the information provided by the Higgs searches will allow very
stringent tests of the SM.  
With its five final state particles, $W+4j$ production gives the opportunity
to study MPI in a
more complex setting than in previous analyses which have typically involved a
combination of two $2 \ra 2$ processes.

The large expected cross section for two jet production suggests that also
Triple Parton Interactions (TPI) could provide a non
negligible contribution\footnote{This issue was raised by the referee,
whose role we gratefully aknowledge.}. If this were the case, it would open the
exciting
possibility of probing TPI, on which we have essentially no information so far.

In \sect{sec:calc} the main features
of the calculation are discussed.
Then we present our results in \sect{sec:MPIvsTT} and  \sect{sec:MPI_in_W4j}.
Finally we summarize the main points of our discussion.

\section{Calculation}
\label{sec:calc}

If no $b$--tagging is assumed the  MPI processes which provide a background to
$t \bar{t}$ and more generally contribute to $Wjjjj$ through Double
Parton Interactions (DPI) are 

\bi
\item $jj \otimes jjW$
\item $jjj \otimes jW$
\item $jjjj \otimes W$
\ei

where the symbol $\otimes$ stands for the combination of one event for each of the two
final states it connects.

The cross section for DPI has been estimated as
\be
\label{eq:sigma_2}
    \sigma =  \sigma_1 \cdot \sigma_2/\sigma_{eff}
\ee
where $\sigma_1 ,\sigma_2$ are the cross sections of the two contributing
reactions. The customary symmetry factor, which is equal to two if the
two reactions are indistinguishable and equal to one when they are different
is always one in the present case. We have used
$\sigma_{eff}=14.5$ mb as measured by CDF \cite{CDF_MPI} at the Tevatron. In 
Ref.\cite{Treleani:2007gi} Treleani argues that a more appropriate value at
$\sqrt{s}= 1.8$ TeV is 10 mb which translates at the LHC into  
$\sigma_{eff}^{LHC}=12$ mb. Since $\sigma_{eff}$ appears as an overall factor
in our results it is easy to take into account the smaller value advocated in 
\cite{Treleani:2007gi}.

The only TPI process contributing to $Wjjjj$ is

\bi
\item $jj \otimes jj \otimes W$.
\ei
It is worth mentioning that this is probably the simplest process and the one
with the largest rate which can give access to Triple Parton Interactions at the
LHC, since it involves two instances of two jet production and a Drell-Yan
interaction which allows to separate this kind of events from the multiple jet
background generated by QCD.
 
The cross section for TPI, under the same hypotheses which lead to
\eqn{eq:sigma_2}, can be expressed as:
\be
\label{eq:sigma_3}
 \sigma =  \sigma_1 \cdot \sigma_2 \cdot \sigma_3/
 \left( \sigma^{\prime }_{eff} \right)^2 /k
\ee

where $k$ is a symmetry factor.
$\sigma^{\prime}_{eff}$ has not been measured, and in principle
it could be different from $\sigma_{eff}$. However, in the
absence of actual data, we will assume $\sigma^{\prime}_{eff} = \sigma_{eff}$
and since two of the reactions in \eqn{eq:sigma_3} are indistinguishable we
will take $k=2$.
In the following we will keep the TPI contribution, which is affected by larger
uncertainties, separated from the DPI predictions which are based on firmer
ground.

Three perturbative orders contribute to $4j + \ell\nu$ at the LHC through
Single Parton Interactions.
The $\ordEW$ and $\ordQCD$ samples have been generated with \Phantom
\cite{PhantomPaper,method,phact}, while the $\ordQCDsq$ sample
has been produced with \MadEvent \cite{MadeventPaper}.
All reactions contributing to MPI have been generated with \MadEvent .
Both programs generate events 
in the Les Houches Accord File Format \cite{LHAFF}.
In all samples full  matrix elements, without any
production times decay approximation, have been used.

The $\ordQCD$ contribution is dominated by
$t\overline{t}$ production. It includes the two main mechanism which yield 
top-antitop pairs, that is $gg \ra t\overline{t}$ and
$q\overline{q} \ra t\overline{t}$. A small, purely electroweak contribution to  
$q\overline{q} \ra t\overline{t}$ processes is included in the $\ordEW$ event set.

All samples have been generated with the following cuts:

\bea
\label{eq:cuts}
& p_{T_j} \geq 30~{\rm GeV} \, , \; \; |\eta_j| \leq 5.0 \, , 
\nonumber \\
& p_{T_\ell} \geq 20~{\rm GeV} \, ,\; \;
|\eta_{\ell}| \leq 3.0 \, , \\
& M_{jj} \geq 60 \, {\rm GeV}  \nonumber 
\eea

where $j= u,\bar{u},d,\bar{d},s,\bar{s},c,\bar{c},b,\bar{b},g$.

The relatively high
$p_{T_j}$ threshold ensures that the processes we are interested in can be
described by perturbative QCD and that our results will not be sensitive to the
details of the low $p_{T}$ underlying event.

\begin{table}[h!tb]
\label{Xsection:MPIgencuts}
\vspace{0.15in}
\begin{center}
\begin{tabular}{|c|c|c|}
\hline
Process &  Cross section  & Combined\\
\hline
 $jj$  &  1.44e8 pb & \multirow{2}{13mm}{4.03 pb}\\
\cline{1-2}
 $jj (\mu^-\bar{\nu_\mu} + \mu^+\nu_\mu)$  &  6.54e2 pb & \\
\hline
 $jjj$ &   7.64e6 pb & \multirow{2}{13mm}{0.68 pb} \\
\cline{1-2}
 $j (\mu^-\bar{\nu_\mu} + \mu^+\nu_\mu)  $ &   1.82e3 pb & \\
\hline
 $jjjj  $ & 1.16e6 pb  & \multirow{2}{13mm}{0.88 pb} \\
\cline{1-2}
 $\mu^-\bar{\nu_\mu} + \mu^+\nu_\mu  $ &     1.09e4 pb & \\
\hline
\end{tabular}
\end{center}
\caption{
Cross sections for the processes which contribute to $4j + \ell\nu$
through DPI. The selection cuts are given in 
\eqn{eq:cuts}. Notice that the combined cross section corresponds to
$\sigma_1\cdot\sigma_2 / \sigma_{eff}$ only for the $jjjj \otimes W$ case. In
all other cases there is a reduction due to the requirement of a minimum
invariant mass for all jet pairs since additional pairs are formed 
when the two events are superimposed.
}
\end{table}

\begin{table}[h!tb]
\label{Xsection:MPIgencuts3}
\vspace{0.15in}
\begin{center}
\begin{tabular}{|c|c|c|}
\hline
Process &  Cross section  & Combined\\
\hline
 $jj$  &  1.44e8 pb & \multirow{3}{13mm}{0.27 pb}\\
\cline{1-2}
 $jj$  &  1.44e8 pb & \\
\cline{1-2}
 $\mu^-\bar{\nu_\mu} + \mu^+\nu_\mu  $ &     1.09e4 pb & \\
\hline
\end{tabular}
\end{center}
\caption{
Cross sections for the processes which contribute to $4j + \ell\nu$
through TPI. The selection cuts are given in 
\eqn{eq:cuts}. 
}
\end{table}

\begin{figure}[h]
\begin{center}
%\mbox{\epsfig{file=Mjjj_log_extra.eps,width=12.cm}}
\mbox{\includegraphics*[width=12.cm]{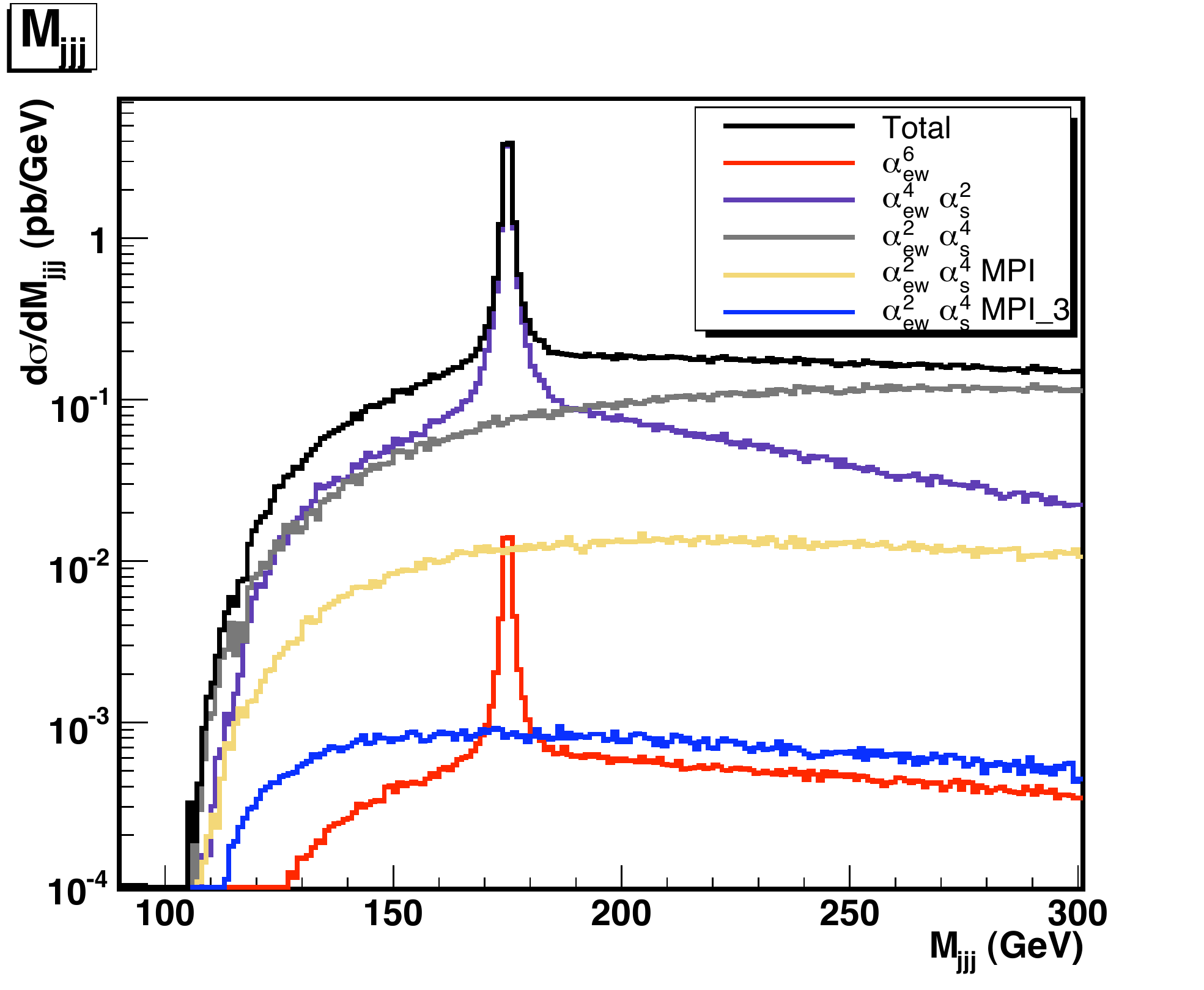}}
\caption {}
$M_{jjj}$ distribution for the different contributions and for their sum.
Cuts as in \eqn{eq:cuts} and \eqn{eq:cuts_iso}.
\label{Mjjj:noflav}
\end{center}
\end{figure}

We have taken $M_t = 175$ GeV.

The cross sections for the reactions which enter the MPI sample are
shown in
%\tbn{Xsection:MPIgencuts}
Tab.1 and 
%\tbn{Xsection:MPIgencuts3}
Tab.2
for DPI and TPI respectively.
We have combined at random one event from each of the reactions which together
produce the  $jjjj (\mu^-\bar{\nu_\mu} + \mu^+\nu_\mu)$ final state through MPI,
and have required that each
pair of colored partons in the final state satisfy $M_{jj} \geq 60 \, {\rm GeV}$.
This implies that the combined cross section corresponds to
to the product of the separate cross sections divided by the appropriate
power of $\sigma_{eff}$ only for the $jjjj \otimes W$ case. In
all other cases there is a reduction due to the requirement of a minimum
invariant mass for all jet pairs since additional pairs are formed 
when the two events are superimposed.
The largest contribution is given by
processes in which the $W$ boson is produced in association with two jets
in one interaction and the
other two jets are produced in the second one. As a consequence, as in
the case of $\gamma+3j$ studied by CDF \cite{CDF_MPI} most of the events contain
a pair of energetic jets with balancing transverse momentum. The next largest
contribution is due to Drell-Yan processes combined with four jet events.
The smallest, but still sizable, DPI contribution is given by
processes in which the $W$ boson is produced in association with one jet,
which balances the $W$ transverse momentum, and the
other three jets are produced in the second interaction.
The cross section for TPI is 0.27 pb, about 5\% of all MPI processes. 

We work at parton level with no showering and hadronization.
Color correlations between the two scatterings have been ignored. They are known
to be important at particle level \cite{Field} but are totally
irrelevant at the generator level we are considering in this paper.
The two jets with the largest and smallest rapidity are identified as forward and
backward jet respectively.
The two intermediate jets will be referred to as central jets in the following.

The neutrino momentum is reconstructed according to the usual prescription,
requiring the 
invariant mass of the $\ell \nu$ pair to be equal to the $W$ boson nominal mass,
\begin{equation}
\label{eq:nu_reco_equation}
(p^{\ell}+p^{\nu})^2 = M_W^2 ,
\end{equation}
in order to determine the longitudinal component of the neutrino momentum.
This equation has two solutions,
\begin{equation}
\label{eq:nu_reco}
p_{z}^{\nu} = \frac{\alpha p_z^\ell \pm \sqrt{\alpha^2 p_z^{\ell 2} - 
 (E^{\ell 2} - p_z^{\ell 2})(E^{\ell 2} p_T^{\nu 2} - \alpha^2)}}
  {E^{\ell 2} - p_z^{\ell 2}}  \; ,
\end{equation}
where
\begin{equation}
\alpha = \frac{M_W^2}{2} + p_x^{\ell}p_x^{\nu} + p_y^{\ell}p_y^{\nu}  \; .
\end{equation}

If the discriminant of Eq.(\ref{eq:nu_reco}) is negative, which happens only if
the actual momenta satisfy $(p^{\ell}+p^{\nu})^2 > M_W^2$,
it is reset to zero.
The corresponding value of $p_{z}^{\nu}$ is adopted.
This value of $p_{z}^{\nu}$ results in the smallest possible value for
the mass of the $\ell \nu$ pair which is compatible with the known components
of $p^{\ell}$ and $p^{\nu}$. The corresponding mass is always larger than $M_W$.
If the determinant is positive and the two solutions
for $p_{z}^{\nu}$ have opposite sign we choose the solution whose sign coincides
with that of $p_{z}^{\ell}$. If they have the same sign we choose the
solution with the smallest $\Delta R$ with the charged lepton.
The reconstructed value is used for computing all physical observables.

All samples have been generated using CTEQ5L \cite{CTEQ5} 
parton distribution functions.
For the $\ordEW$ and $\ordQCD$ samples, generated with \Phantom,
the QCD scale has been taken as
\be
\label{eq:LargeScaleTop}
Q^2 = M_W^2 + p_{Ttop}^2
\ee
if a triplet of final state particles with flavours compatible with deriving
from the decay of a top or antitop quark can be found, while it has been taken
as
\be
\label{eq:LargeScale}
Q^2 = M_W^2 + \frac{1}{6}\,\sum_{i=1}^6 p_{Ti}^2
\ee
in all other cases.
For the $\ordQCDsq$ sample the scale
has been set to $Q^2 = M_Z^2$. This difference in the scales
leads to a definite relative enhancement of
the  $4j\, +  \, W$ background and of the MPI contribution compared to the other
ones. Tests in
comparable reactions have shown an increase of about a factor of 1.5 
for the processes computed at $Q^2 = M_Z^2$ with
respect to the same processes computed with the larger scale
\eqn{eq:LargeScale}.

In our estimates below we have only taken into account the muon
decay of the $W$ boson. The $W\ra e\nu$ channel gives the same result.
The possibility of detecting high $p_T$ taus has been
extensively studied in connection with the discovery of a light Higgs in Vector
Boson Fusion in the
$\tau^+\tau^-$ channel \cite{ATLAS-HinVV} with extremely encouraging results.
Efficiencies of order 50\% have been obtained for the hadronic decays of the
$\tau's$. The expected number of events in the $H\rightarrow  \tau\tau
\rightarrow e\mu + X$ is within a factor of two of the yield from 
$H\rightarrow  W W^* \rightarrow e\mu + X$ for $M_H = 120$ \GeV where the
$\tau\tau$ and $W W^*$ branching ratios of the Higgs boson are very close,
suggesting that also in the leptonic decay channels of the taus the efficiency
is quite high. Therefore we expect the $W \rightarrow \tau \nu$ channel
to increase the detectability of the $Wjjjj$ final state.

\begin{figure}[h]
\begin{center}
%\mbox{\epsfig{file=DeltaEta_jfjb_extra.eps,width=12.cm}}
\mbox{\includegraphics*[width=12.cm]{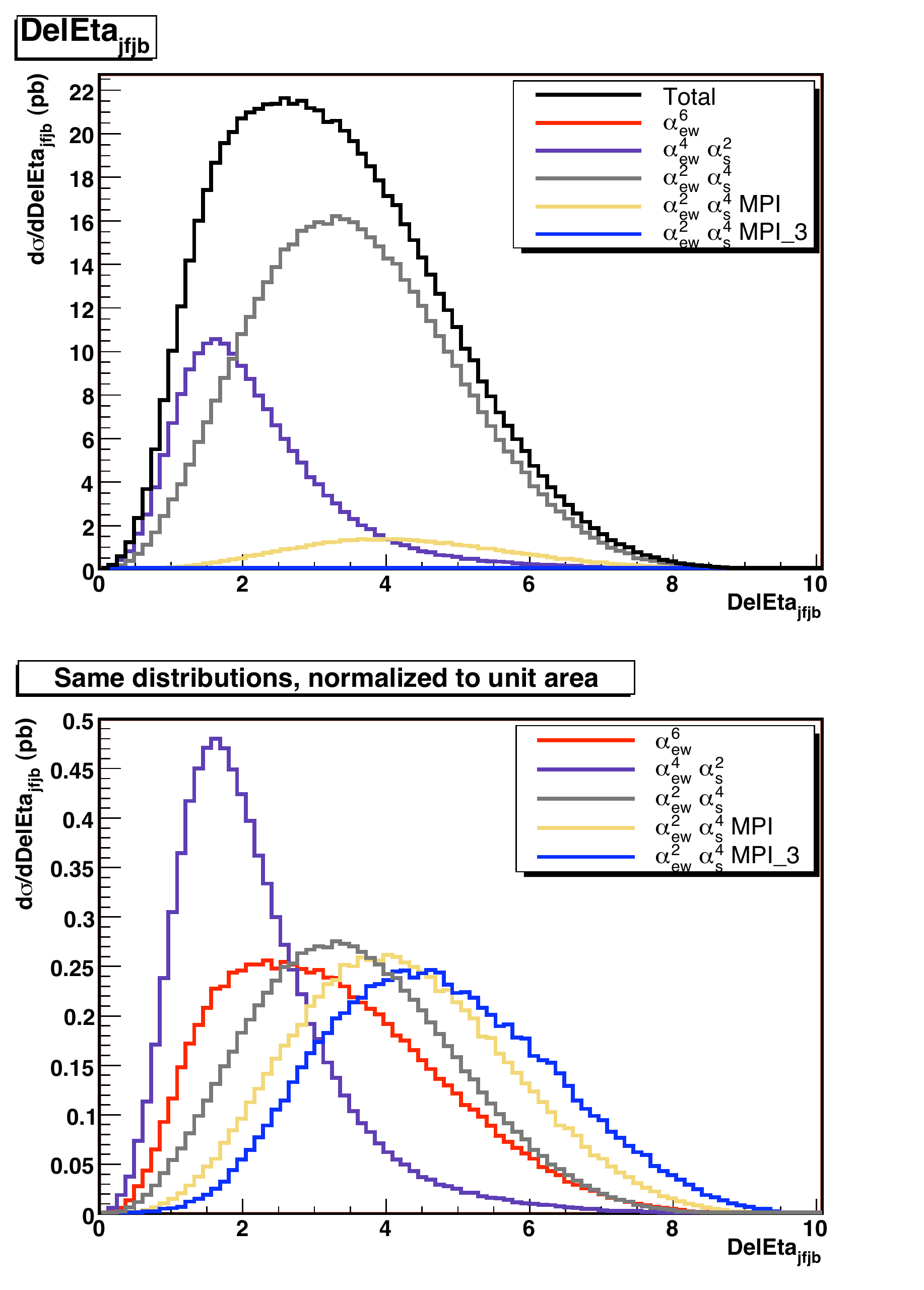}}
\caption {}
$\Delta\eta$ separation between the most forward and most backward jet 
for the different contributions and for their sum.
Cuts as in \eqn{eq:cuts} and \eqn{eq:cuts_iso}.
The curves in the lower plot are normalized to one.
\label{DeltaEta:noflav}
\end{center}
\end{figure}

\section{MPI background to top pair production}
\label{sec:MPIvsTT}

The cross section for Single Particle Interaction processes and Multiple Parton
Interactions contributing to  the  $jjjj (\mu^-\bar{\nu_\mu} + \mu^+\nu_\mu)$ final state,
with the set of cuts in \eqn{eq:cuts}, are shown in the second column of
%\tbn{Xsection:gencuts} 
Tab. 3.
The cross sections in the third column have been obtained with the additional requirements:
\be
\Delta R(jj) > 0.5 \; \;\;\; \;\;
\Delta R(jl^\pm) > 0.5
\label{eq:cuts_iso}
\ee
which ensure that all jet pairs are well separated and that the charged lepton is isolated.

\begin{figure}[h]
\begin{center}
%\mbox{\epsfig{file=Mvis_extra.eps,width=12.cm}}
\mbox{\includegraphics*[width=12.cm]{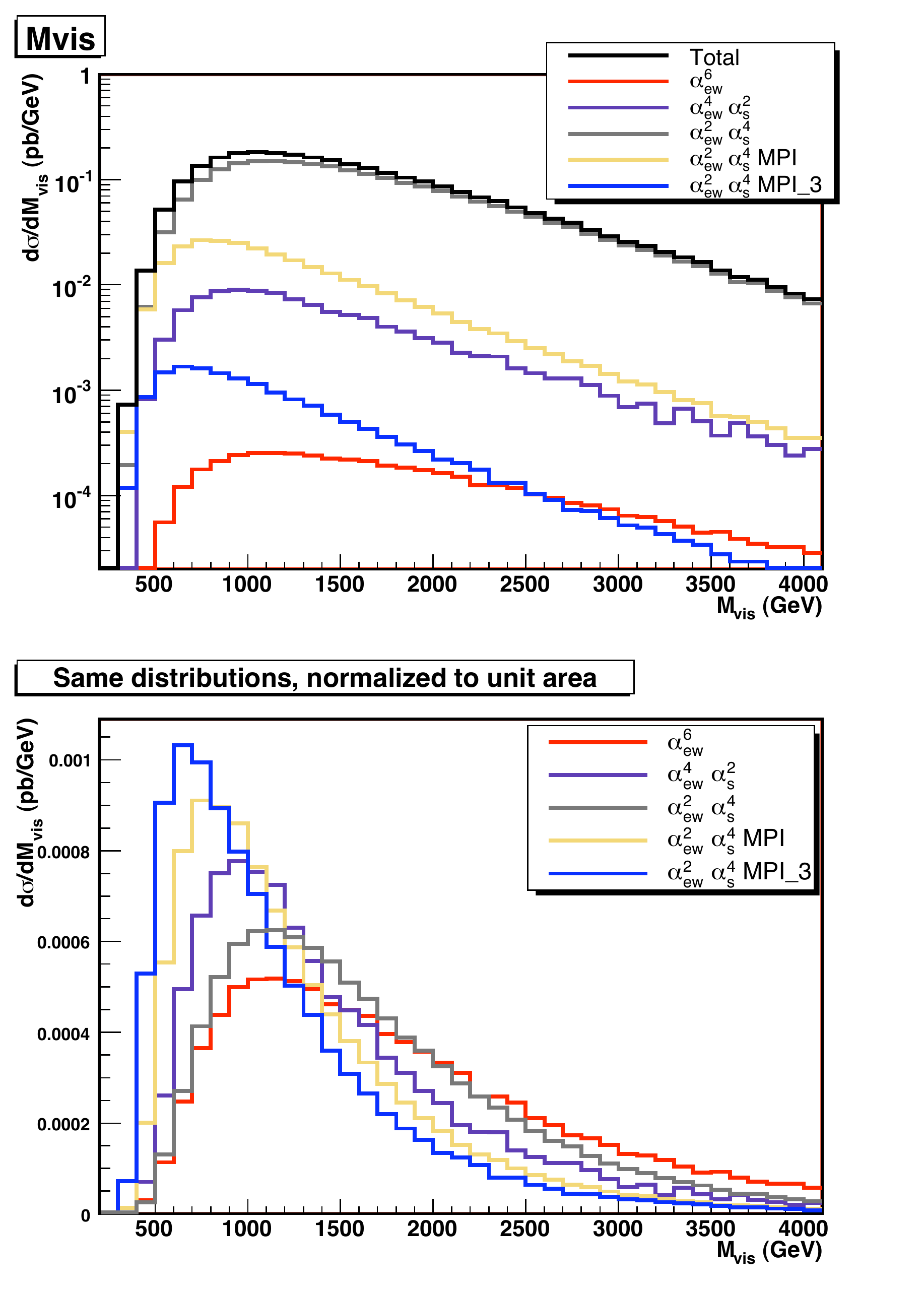}}
\caption{
Visible mass distribution
for the different contributions and for their sum.
Cuts as in \eqn{eq:cuts}, \eqn{eq:cuts_iso} and \eqn{cuts:DeltaEta}.
The curves in the lower plot are normalized to one.
}
\label{Mvis:noflav}
\end{center}
\end{figure}

\begin{table}[h!tb]
\label{Xsection:gencuts}
\vspace{0.15in}
\begin{center}
\begin{tabular}{|c|c|c|}
\hline
Process &  Cross section &  Cross section \\
\hline
$\ordQCD$  &  25.0  pb & 22.0  pb\\
\hline
$\ordQCDsq$  & 64.7  pb & 58.9  pb\\
\hline
$\ordQCDsq_{\mathrm{DPI}}$  & 5.6  pb & 5.3  pb\\
\hline
$\ordQCDsq_{\mathrm{TPI}}$  & 0.27  pb & 0.26  pb\\
\hline
$\ordEW$  & 0.22 pb & 0.20 pb \\
\hline
\end{tabular}
\end{center}
\caption{Cross sections for the processes which contribute to $4j + \ell\nu$.
For the second column the selection cuts are given in \eqn{eq:cuts}.
For the third column the additional requirements  \eqn{eq:cuts_iso} have been
applied.}
\end{table}

In \fig{Mjjj:noflav} we show the invariant mass distribution of the
triplet of jets with the highest
combined transverse momentum which is expected to provide a good
measurement of the top mass in the early phase
of LHC \cite{ref:EarlyMass}, when $b$--tagging might be unavailable.
The cross sections for masses in the range
170~GeV~$<  M_{jjj} <$~180~GeV are presented in
%\tbn{Xsection:peak}
Tab. 3.  This small mass interval has been selected purely for illustrational
purposes. The actual smearing of the observed mass peak will be dominated by the
uncertainty on the jet energy scale.

\begin{table}[h!tb]
\label{Xsection:peak}
\vspace{0.15in}
\begin{center}
\begin{tabular}{|c|c|}
\hline
Process &  Cross section\\
\hline
$\ordQCD$  &  10.8  pb \\
\hline
$\ordQCDsq$  & 0.76  pb \\
\hline
$\ordQCDsq_{\mathrm{DPI}}$  & 0.12  pb \\
\hline
$\ordQCDsq_{\mathrm{TPI}}$  & 0.01  pb \\
\hline
$\ordEW$  & 0.04 pb \\
\hline
\end{tabular}
\end{center}
\caption {Total cross sections in the mass range
170 GeV $<  M_{jjj} <$ 180 GeV . Cuts as in \eqn{eq:cuts} 
and \eqn{eq:cuts_iso}.}
\end{table}

The $W+4j$ $\ordQCDsq$ processes therefore provide a background of about 7\% to top--antitop production 
in the semileptonic channel, while MPI processes provide a 1\% background in
this mass range. While such an increase of the background is unlikely to affect
the mass measurement, it might be relevant for the measurement of the cross
section for $t\bar{t}$ production whose uncertainty is dominated by the
background normalization \cite{ref:TopCrossSEction}.

It is obvious from the results in 
%\tbn{Xsection:peak}
Tab. 3, that once $b$--tagging will be fully operational both the $W+4j$ 
background and the MPI background will be completely negligible since only a
very small fraction of events in these two samples contain $b$ quarks while two
$b$'s are always present in $t \bar{t}$ events.   

\section{Studying MPI in $W+4j$ processes}
\label{sec:MPI_in_W4j}

Any attempt to detect MPI processes in the $jj (\mu^-\bar{\nu_\mu} + \mu^+\nu_\mu)$
channel requires a strong suppression of top--antitop production. For this purpose,
we have required that no jet triplet satisfies
\be
\label{eq:topcut1}
\vert M_{jjj} - M_t \vert < 10 \,{\mathrm GeV}
\ee

\fig{DeltaEta:noflav} shows that MPI events tend to have larger separation in pseudorapidity
between the most forward and most backward jets than $W+4j$ or $t\bar{t}$ production.
Therefore we require:
\be
\label{cuts:DeltaEta}
|\Delta\eta(j_fj_b)| > 3.8
\ee

In a more realistic environment in which additional jets generated by showering cannot be ignored,
one could impose condition (\ref{eq:topcut1}) and (\ref{cuts:DeltaEta})
 on the four most energetic jets in the event.
The cross sections obtained after vetoing top production \eqn{eq:topcut1}
and with the set of 
cuts in \eqn{eq:cuts}, \eqn{eq:cuts_iso} and \eqn{cuts:DeltaEta} are shown
in 
%\tbn{Xsection:MPIvsBKG}
Tab. 5.
Assuming a luminosity of 100 pb$^{-1}$ this corresponds to a statistical
significance of the MPI signal of about 6.1 if we take into acccount both the
DPI and TPI contributions, and of 5.8 if we conservatively consider only DPI
processes.

\begin{figure}[h]
\begin{center}
%\mbox{\epsfig{file=DeltaPhi_jjmax_extra.eps,width=12.cm}}
\mbox{\includegraphics*[width=12.cm]{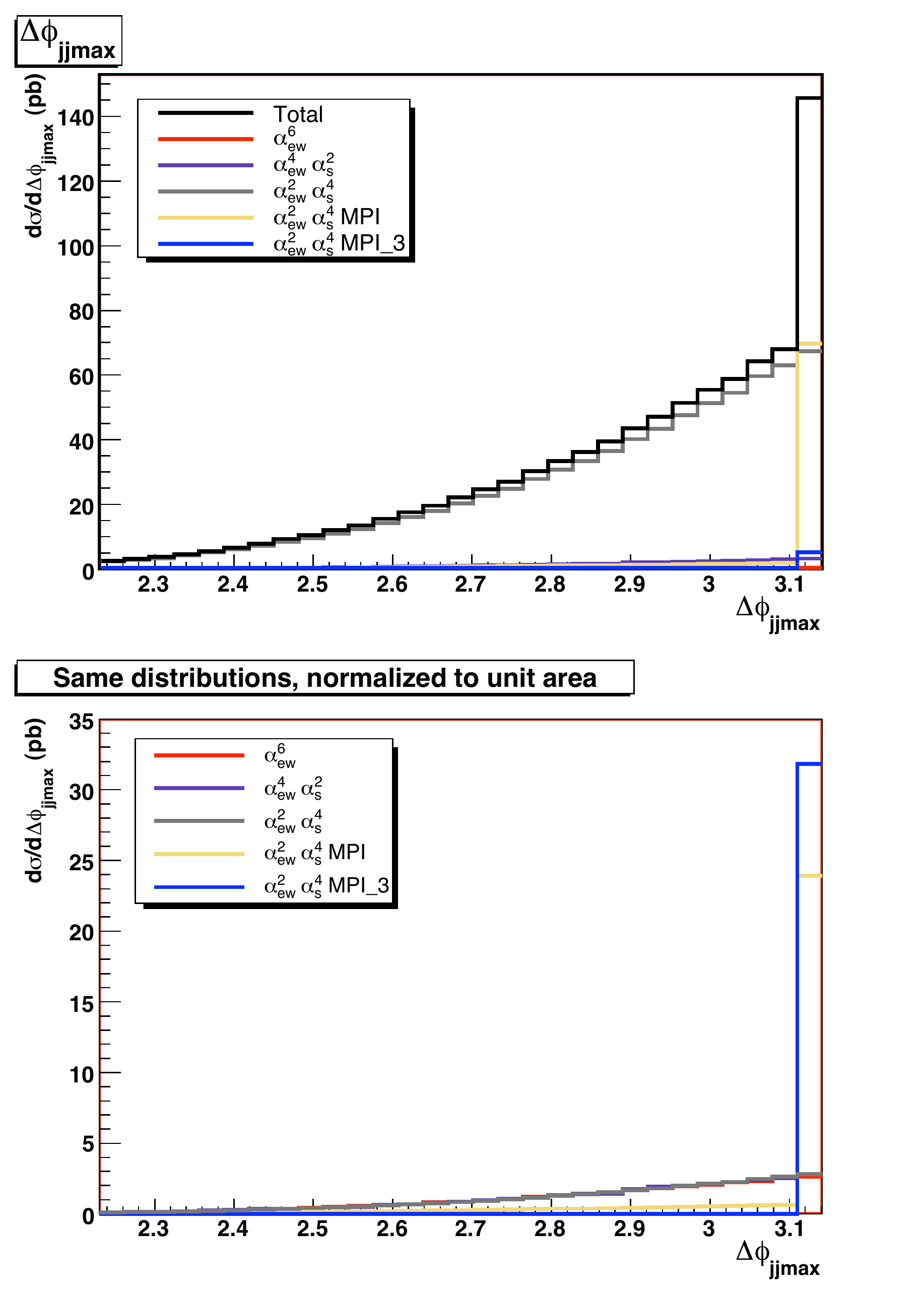}}
\caption{
Largest $\Delta\phi$ separation between jet pairs
for the different contributions and for their sum.
Cuts as in \eqn{eq:cuts}, \eqn{eq:cuts_iso} and \eqn{cuts:DeltaEta}.
The curves in the lower plot are normalized to one.
}
\label{DeltaPhi_jjmax:noflav}
\end{center}
\end{figure}

\begin{table}[h!tb]
\label{Xsection:MPIvsBKG}
\vspace{0.15in}
\begin{center}
\begin{tabular}{|c|c|}
\hline
Process &  Cross section\\
\hline
$\ordQCD$  &  1.16  pb \\
\hline
$\ordQCDsq$  & 24.01  pb \\
\hline
$\ordQCDsq_{\mathrm{DPI}}$  & 2.91  pb \\
\hline
$\ordQCDsq_{\mathrm{TPI}}$  & 0.16  pb \\
\hline
$\ordEW$  & 0.05 pb \\
\hline
\end{tabular}
\end{center}
\caption{Total cross sections after vetoing the top, \eqn{eq:topcut1}.
Cuts as in \eqn{eq:cuts}, \eqn{eq:cuts_iso}
and \eqn{cuts:DeltaEta}.
}
\end{table}

\fig{Mvis:noflav} presents the distribution on the invariant mass of the four
jet plus charged lepton system. It shows that typically MPI events are less
energetic than all other contributions considered in this paper.

\begin{figure}[h]
\begin{center}
\hspace*{-2cm}
%\mbox{\epsfig{file=DeltaPhi_j1j2_extra.eps,width=8.3cm}}
\includegraphics*[width=8.3cm]{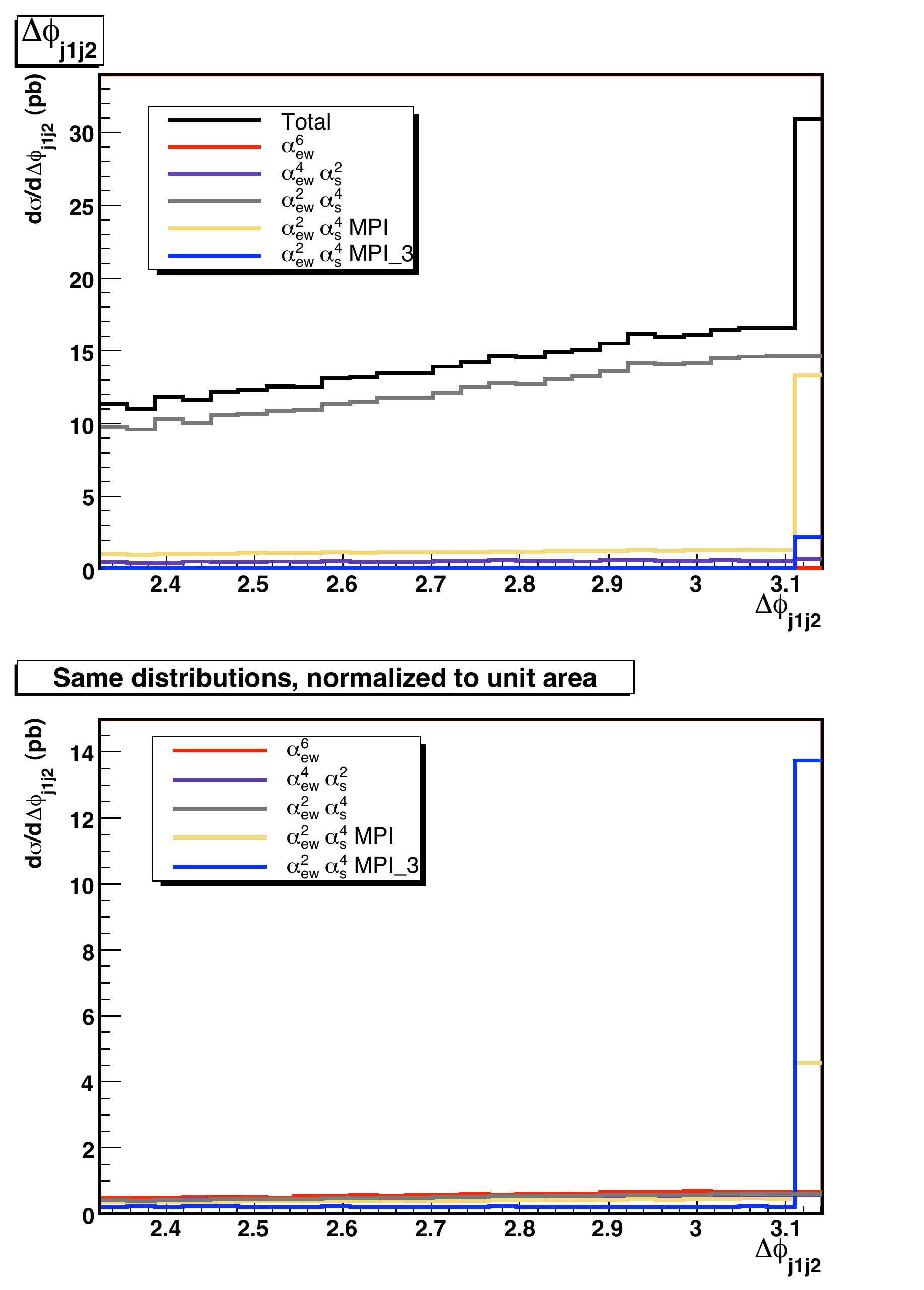}
\hspace*{-0.6cm}
%\mbox{\epsfig{file=DeltaPhi_j3j4_extra.eps,width=8.3cm}}
\includegraphics*[width=8.3cm]{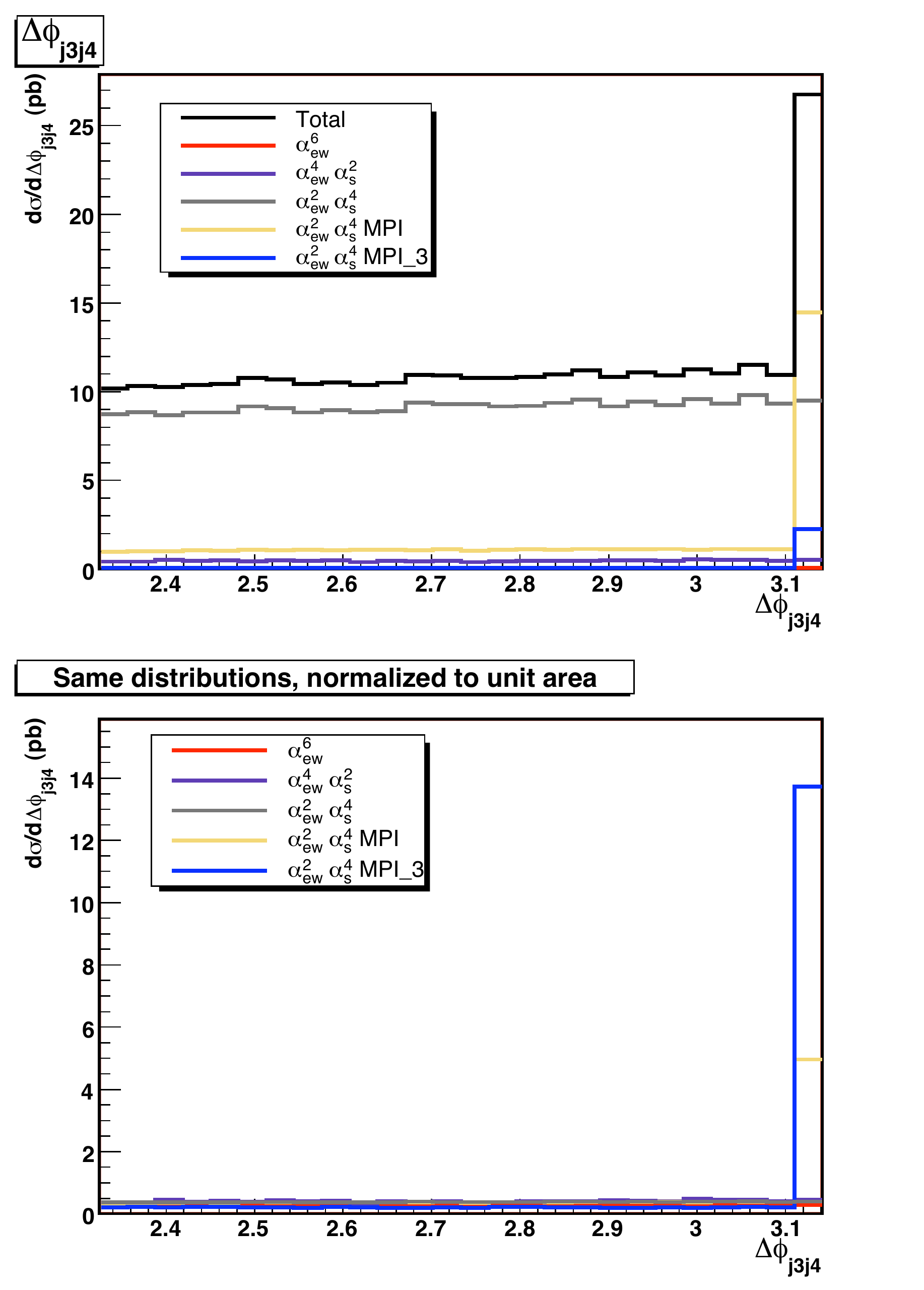}
\hspace*{-2cm}
\caption{
$\Delta\phi$ separation between the two most energetic jets (on the left)
and between the two least energetic among
the four jets (on the right)
for the different contributions and for their sum.
Cuts as in \eqn{eq:cuts}, \eqn{eq:cuts_iso} and \eqn{cuts:DeltaEta}.
The curves in the lower plot are normalized to one.
}
\label{DeltaPhi_jj:noflav}
\end{center}
\end{figure}

In \fig{DeltaPhi_jjmax:noflav} we present the distribution of the
largest $\Delta\phi$ separation between all jet pairs.
\fig{DeltaPhi_jjmax:noflav} confirms that MPI processes
leading to $W+4j$
events are characterized by the presence of two jets which are back to back in
the transverse plane. 
The $W+4j$ $\ordQCDsq$ contribution displays a much milder increase in
the back to back region. All other contributions are negligible.

The expected $\Delta\phi$ resolution is of the order of
a few degrees for both ATLAS \cite{ATLAS-TDR} and CMS \cite{CMS-TDR} for jets
with transverse energy above 50 GeV. This resolution is comparable
to the width of the bins in \fig{DeltaPhi_jjmax:noflav}.
We have examined the $\Delta\phi$  separation among pairs of jets ordered in
energy, $E_{j_i} > E_{j_{i+1}}$. No clear pattern has emerged.
In \fig{DeltaPhi_jj:noflav} we show the 
$\Delta\phi$ separation between the two most  energetic jets, on the left,
and of the two least energetic ones, on the right.
As might have been guessed by the visible mass distribution in
\fig{Mvis:noflav} the ratio between the MPI signal at $\Delta\phi = \pi$ 
and the $W+4j$ background is somewhat larger for softer jet pairs than for
harder ones. It has proved impossible to clearly associate the two balancing jets with either the most
forward/backward pair or with the central jets.

In conclusion, it appears  quite feasible to achieve a good signal to background ratio,
close to 1/1, for Multiple Interactions  Processes compared to Single
Interaction ones by selecting events with two jets with 
$180^\circ$ separation in the transverse plane. It has not been possible to
characterize further this pair of jets either through their energy or angular
ordering.

\begin{figure}[h]
\begin{center}
%\mbox{\epsfig{file=DeltaPhi_comp.eps,width=12.cm}}
\mbox{\includegraphics*[width=12.cm]{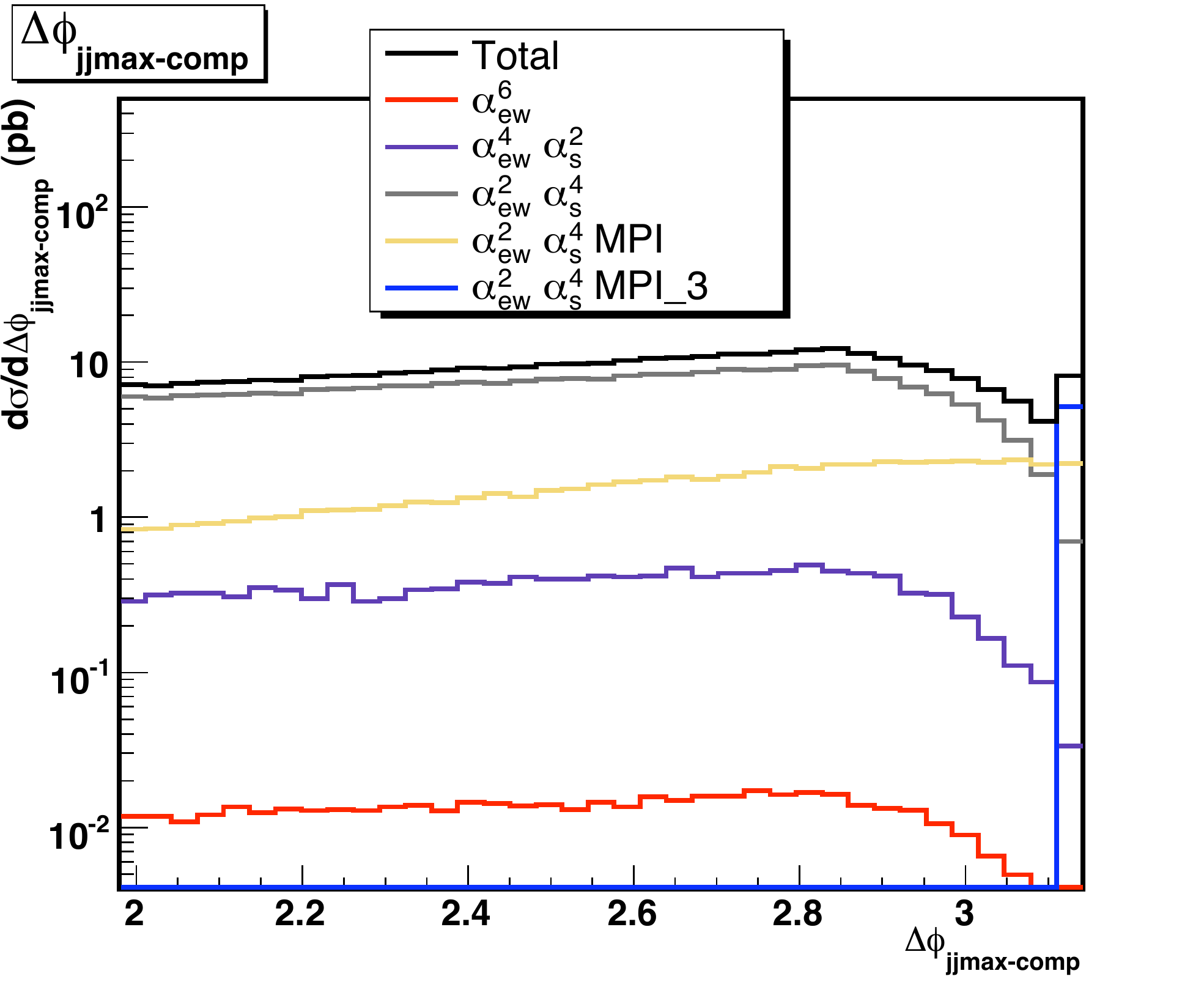}}
\caption{
$\Delta\phi$ separation between the two jets which do not belong to the pair
with the largest $\Delta\phi$ in the event.
Cuts as in \eqn{eq:cuts}, \eqn{eq:cuts_iso}, \eqn{cuts:DeltaEta}
and \eqn{cuts:DeltaPhi}.
}
\label{DeltaPhi_jjmax_comp}
\end{center}
\end{figure}

\begin{figure}[h]
\begin{center}
%\mbox{\epsfig{file=pTlv.eps,width=12.cm}}
\mbox{\includegraphics*[width=12.cm]{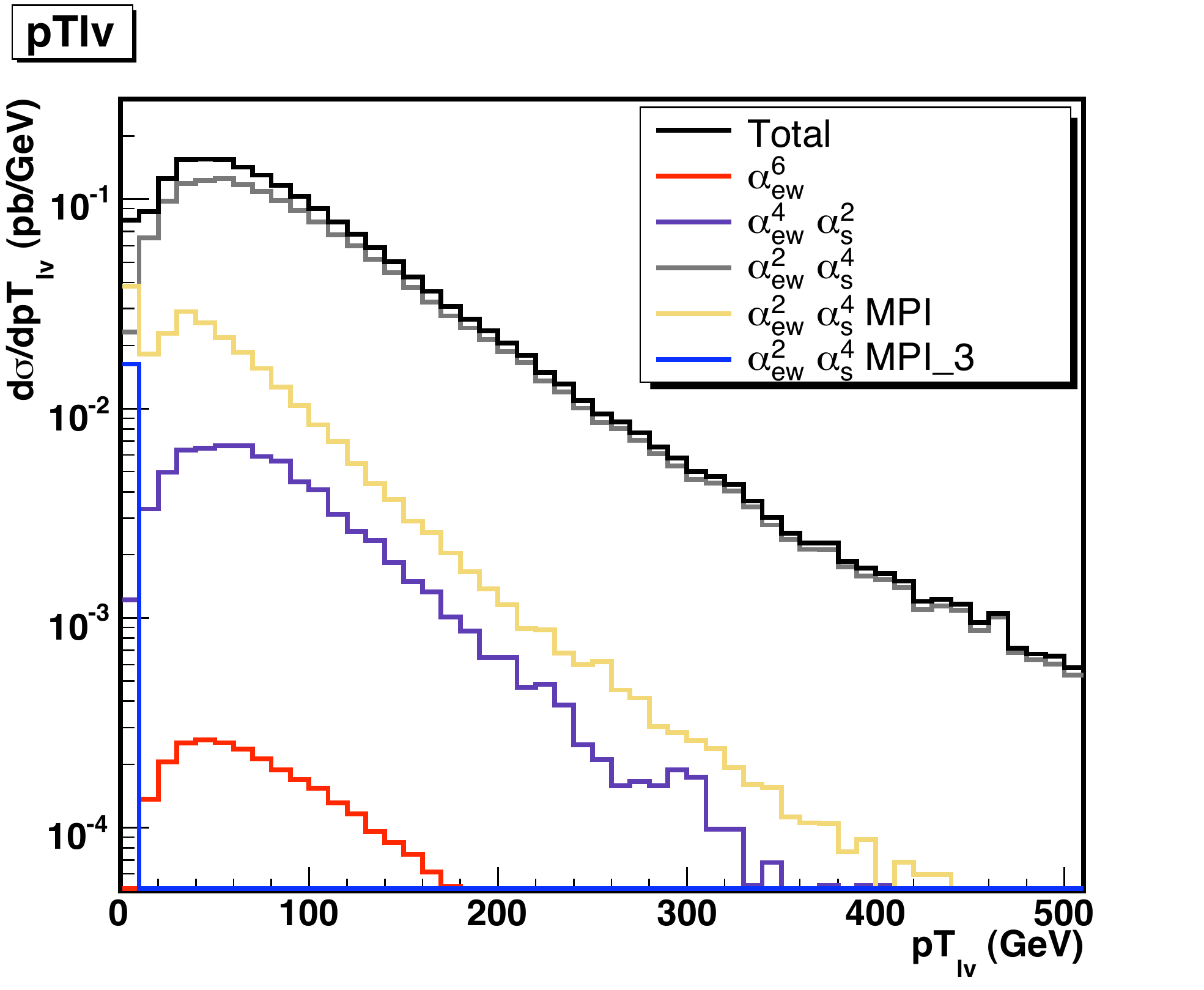}}
\caption{
Distribution of the transverse momentum of the $l\nu$ system.
Cuts as in \eqn{eq:cuts}, \eqn{eq:cuts_iso}, \eqn{cuts:DeltaEta}
and \eqn{cuts:DeltaPhi}.
}
\label{pTlv}
\end{center}
\end{figure}

Let us now discuss Triple Parton Interactions in more detail. The obvious
traits which characterize these events are the presence of two pairs of jets
which balance in transverse momentum and of one $W$ produced by a Drell-Yan
interaction in which, to lowest order, the transverse momentum of the charged
lepton is equal and opposite to the missing $pT$.
While the first feature is not typically found in DPI, $W$ bosons of Drell-Yan
origin are prsent in $jjjj \otimes W$ events which account for about 15\%
of DPI. This is illustrated in \fig{DeltaPhi_jjmax_comp} and \fig{pTlv}.
 For these two plots we have only considered events in
which the maximum  $\Delta\phi$ among jets is in the interval: 

\be
\label{cuts:DeltaPhi}
|\Delta\phi(jj)_{max}| > 0.9\cdot\pi
\ee
The corresponding cross sections are shown in
%\tbn{Xsection:MPI3}
Tab. 6. The rate decrease is of the order of 30\% for Single Parton Interactions
and essentially negligible for MPI processes. The corresponding rates at the LHC
are sizable.
Even at low luminosity, L = 30 fb$^{-1}$, about 5000 TPI events per year are
expected.  

\begin{table}[h!tb]
\label{Xsection:MPI3}
\vspace{0.15in}
\begin{center}
\begin{tabular}{|c|c|}
\hline
Process &  Cross section\\
\hline
$\ordQCD$  &  0.75  pb \\
\hline
$\ordQCDsq$  & 15.61  pb \\
\hline
$\ordQCDsq_{\mathrm{DPI}}$  & 2.61  pb \\
\hline
$\ordQCDsq_{\mathrm{TPI}}$  & 0.16  pb \\
\hline
$\ordEW$  & 0.03 pb \\
\hline
\end{tabular}
\end{center}
\caption{Total cross sections with selection cuts in
Eqs. \ref{eq:cuts},~\ref{eq:cuts_iso},~\ref{eq:topcut1},~\ref{cuts:DeltaEta}
and \ref{cuts:DeltaPhi}.
}
\end{table}

\fig{DeltaPhi_jjmax_comp} shows the angular separation in the tranverse plane
between the two jets which do not belong to the pair with the largest
$\Delta\phi$ in the event. 
The TPI contribution is concentrated at 180$^\circ$
while all other distributions are rather flat in that region. With the
normalization $\sigma^{\prime}_{eff} = \sigma_{eff}$ in \eqn{eq:sigma_3}, 
TPI give  the largest contribution in the bin at $\Delta\phi = \pi$,
amounting to more than 50\% of the total. 

\fig{pTlv} suggests that the presence of a charged lepton whose transverse
momentum balances the missing $pT$ is of limited use in separating TPI events
from their background.

Because of the lack of information concerning the rate of Triple Parton
Interactions, it is impossible to draw any firm conclusion from our preliminary
analysis; \fig{DeltaPhi_jjmax_comp} however suggests that indeed it might be
possible to investigate TPI at the LHC exploiting the angular distribution of
pairs of jets.

\section{Conclusions}
\label{sec:conclusions}

In this paper we have estimated for the first time the background provided by 
Multiple Parton Interactions to top--antitop production in the semileptonic
channel at the LHC. We have concentrated on the early phase of data taking in
which the mass will be measured from the invariant mass of jet triplets without
resorting to $b$--tagging. The MPI background is about 1\% in the mass region
$M_{top}\pm 5$ GeV, to be compared with a background of about 7\% from
$W+4j$ via Single Parton Interactions.

The MPI contribution to $W+4j$ is dominated by events with two jets with
balancing transverse momentum. Both ATLAS and CMS have good resolution in the
polar angle $\phi$, and it looks possible to extract the MPI signal in this
channel. Comparison with other reactions in which MPI processes can be measured
should allow detailed studies of the flavour and fractional momentum dependence
of Multiple Parton Interactions.
 
Our preliminary analysis suggests that it might be
possible to investigate TPI at the LHC using the $jj \otimes jj \otimes W$
channel, which is the simplest process and the one
with the largest rate which can give access to Triple Parton Interactions,
exploiting the angular distribution of pairs of jets.

\section *{Acknowledgments}
We gratefully acknowledge several illuminating
discussions with Roberto Chierici and Paolo
Bartalini.\\
This work has been supported by MIUR under contract 2006020509\_004 and by the
European Community's Marie-Curie Research 
Training Network under contract MRTN-CT-2006-035505 `Tools and Precision
Calculations for Physics Discoveries at Colliders'

%\begin{thebibliography}{999}

\end{document}